\documentclass{article}
\usepackage{spconf,amsmath,graphicx,cite}
\usepackage{xcolor}
\usepackage{booktabs}

\newcommand\blfootnote[1]{%
  \begingroup
  \renewcommand\thefootnote{}\footnote{#1}%
  \addtocounter{footnote}{-1}%
  \endgroup
}

\title{Discrete Unit based
Masking for Improving Disentanglement in Voice Conversion}
\name{Philip H. Lee*, Ismail Rasim Ulgen*, Berrak Sisman}
\address{Speech \& Machine Learning (SML) Lab, The University of Texas at Dallas, USA}
\begin{document}
%
\maketitle
\begin{abstract}





Voice conversion (VC) aims to modify the speaker’s identity while preserving the linguistic content. Commonly, VC methods use an encoder-decoder architecture, where disentangling the speaker's identity from linguistic information is crucial. However, the disentanglement approaches used in these methods are limited as the speaker features depend on the phonetic content of the utterance, compromising disentanglement. This dependency is amplified with attention-based methods. To address this, we introduce a novel masking mechanism in the input before speaker encoding, masking certain discrete speech units that correspond highly with phoneme classes. Our work aims to reduce the phonetic dependency of speaker features by restricting access to some phonetic information. Furthermore, since our approach is at the input level, it is applicable to any encoder-decoder based VC framework. Our approach improves disentanglement and conversion performance across multiple VC methods, showing significant effectiveness, particularly in attention-based method, with 44\% relative improvement in objective intelligibility.

\end{abstract}
\section{Introduction}
\label{sec:intro}

\begin{figure}[!t]
\centering

    \scalebox{0.65}{\includegraphics{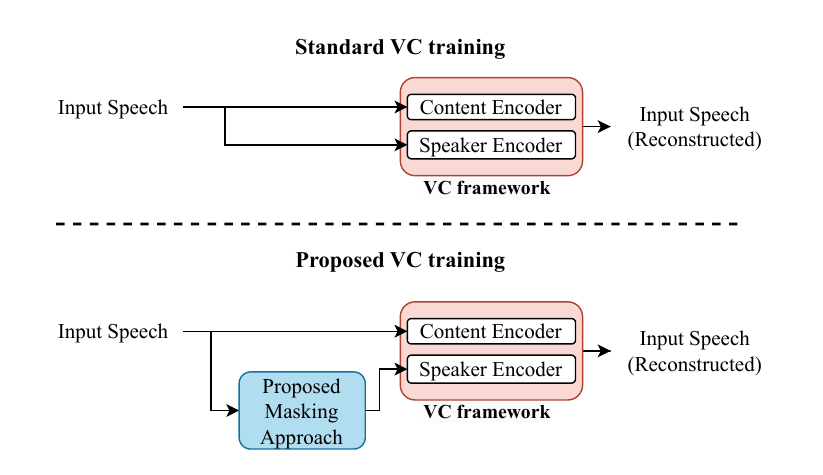}}

    \caption{Standard encoder-decoder based VC training vs VC training with the proposed masking approach}

    \label{fig:concept}
\vspace{-5mm}
\end{figure}

 
Speaker identity plays a crucial role in voice conversion (VC) as it enables the synthesized speech to be personalized for various applications. Historically, early VC methods relied on parametric and non-parametric approaches of statistical learning \cite{sisman2020overview}. In the deep learning era, methods initially focused on VC with parallel training data \cite{parallel-dl1,parallel-dl,parallel-dl2}. However, the introduction of advanced networks such as generative adversarial networks (GANs) \cite{cycle-gan-vc,Kameoka2020IEEETrans_StarGAN-VC} and variational autoencoders (VAEs)\cite{qian2019autovc} 
opened up ways to train VC architectures without any parallel training data. Recently, encoder-decoder-based methods have been the state-of-the-art (SOTA) and most-adopted approach \cite{adain-vc,wang2021vqmivc,qian2019autovc,again-vc} (as reported in Figure \ref{fig:concept}), allowing more breakthroughs in voice quality and speaker similarity in VC.
\blfootnote{*Equal contribution} 
\blfootnote{\textbf{Speech samples:} https://conv-synth.github.io/unit-masked-VC/} 






Zero-shot VC techniques aim to transform source speech into the voice of a target speaker using a limited amount of reference speech from that target, thereby allowing the usability of VC methods across various scenarios. Popular zero-shot VC methods have an encoder-decoder architecture \cite{wang2021vqmivc,again-vc,adain-vc} where speech is encoded to linguistic and speaker features, which are then decoded back to speech. Voice conversion is performed by using source linguistic information and target speaker information during decoding. The most crucial points in encoder-decoder-based approaches are the disentanglement and descriptive power of linguistic and speaker features to synthesize high-quality speech while controlling the speaker identity.




There are different disentanglement approaches adopted by the SOTA VC methods \cite{qian2019autovc,adain-vc,wang2021vqmivc,again-vc,du22c_interspeech}. In \cite{qian2019autovc}, the authors attain disentanglement by limiting the bottleneck feature dimension of linguistic features. In \cite{adain-vc}, the authors introduce adaptive instance normalization for disentanglement. In \cite{wang2021vqmivc}, to disentangle encoded features, the authors impose a mutual information estimation and minimization objective on encoded features. All these disentanglement approaches are enforced at the feature level and seem to be effective. However, they have limitations in terms of disentanglement, and we suggest that all these feature-level disentanglement approaches can be supported by additional input-level approaches.

As representation ability is another significant challenge in VC, attention-based feature extraction methods are introduced to obtain more fine-grained, detailed speaker identity information from input speech. In \cite{Park2023TriAANVCTA}, the authors proposed TriAAN-VC where they modify the instance normalization by introducing attention-based statistics calculation. In \cite{Li2023SEFVCSE}, the authors extract speaker identity information through the cross-attention between discrete semantic tokens and frame-level speaker features. However, we suggest that having more detailed information through attention is likely to compromise disentanglement, and this approach needs a remedy.


It is shown that speaker identity information is connected with the phonetics of the speech, and speaker features tend to contain phonetic information and are dependent on the phonetic structure of the input speech \cite{Raj2019ProbingTI,Hong2023DecompositionAR,Shon2018FrameLevelSE}. We believe this phenomenon is harmful to zero-shot, encoder-decoder-based VC methods as this dependency is a form of entanglement. In the most adopted approach, the network is trained with a reconstruction objective where speaker features are extracted from the same phonetic structure of the source speech. However, in the inference scenario, the reference speech from the target speaker might not have the same phonetic structure as the source speech, degrading the ability to extract speaker information and potentially causing overfitting.


In this paper, we propose a very straightforward yet effective way to reduce the phonetic dependency of speaker features and improve disentanglement in VC as reported in Figure \ref{fig:concept} conceptually. In VC training, we apply masking to the utterance before feeding it into the speaker encoder such that we mask all occurrences of randomly selected discrete speech units in the speech that are highly correlated with phonemes. We apply our approach to multiple VC frameworks to show that it applies to different VC methods regardless of the framework. We show that our method is effective in both VC frameworks, especially with remarkable improvements in the attention-based framework, proving to be a remedy for the potential phonetic dependency of those methods.

\vspace{-3mm}
\section{Related Work}
\vspace*{-2mm}



Masking on input is applied in various ways for different purposes in VC \cite{kaneko2021MaskCycleGAN-VC,seq2seq-vc,choi23d_interspeech,terashima2022cross,perturb}. In \cite{kaneko2021MaskCycleGAN-VC}, a GAN-based method, the authors applied time-masking to the input to enforce the model to predict the missing frames, which resulted in improved VC performance. In \cite{seq2seq-vc}, the authors applied multiple maskings, including time and frequency masking, to augment the limited parallel data. More recently, in \cite{choi23d_interspeech}, the authors applied masking to the prior of the diffusion model to improve robustness. Most similar to our work in terms of motivation, in \cite{terashima2022cross,perturb}, the authors perturb the pitch of input speech before feeding it to the linguistic encoder to prohibit its access to speaker information. This method has been proven to be very effective in improving disentanglement in both works. 
To our knowledge, this is the first work to address the phonetic dependency of speaker features and focus on information perturbation specifically designed for speaker encoders in VC methods.

\begin{figure*}[t]

    \centering
    \scalebox{0.45}
    {\includegraphics{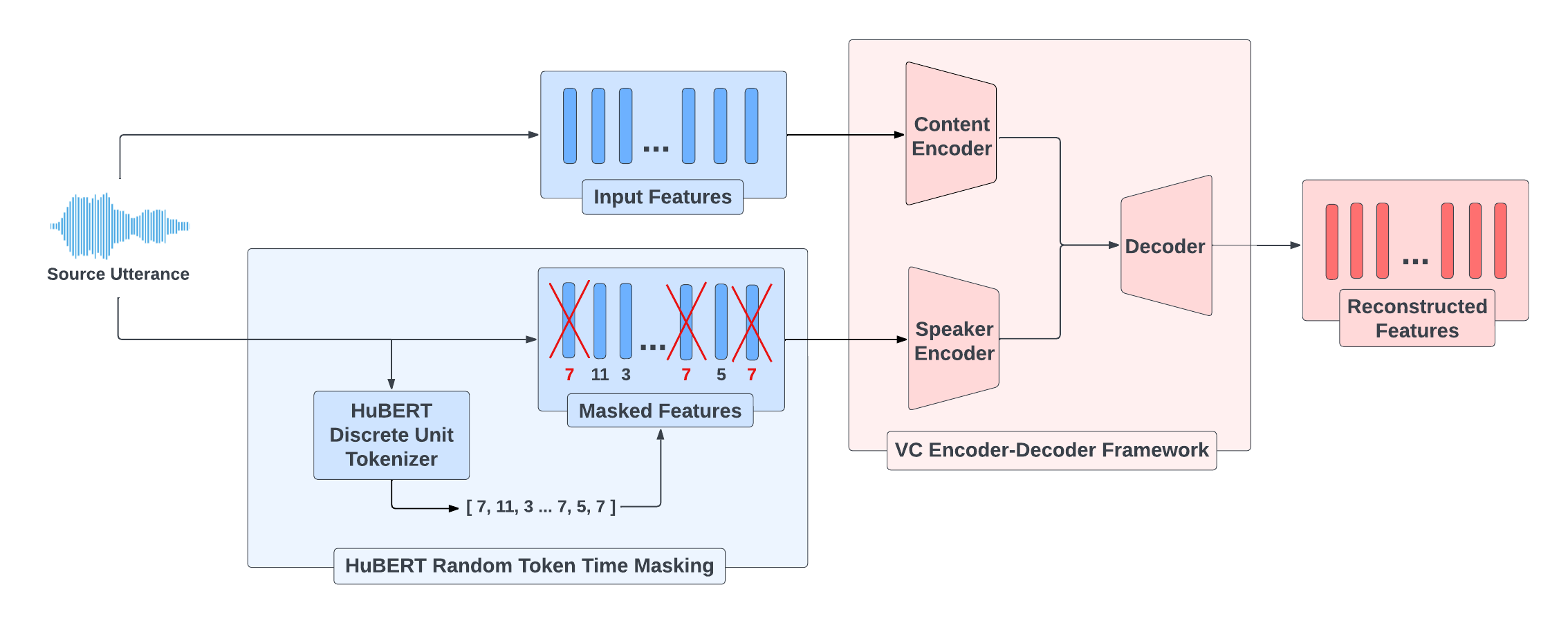}}
    \caption{Proposed training for encoder-decoder based VC frameworks}

    \label{fig:framework}

\end{figure*}

\section{Discrete Unit based Masking for Voice Conversion}


We propose a novel masking approach specifically designed for encoder-decoder based, zero-shot voice conversion. Our masking approach focuses on disrupting the phonetic content of the input utterance before feeding it to the speaker encoder. We aim to reduce the phonetic dependency of speaker features and improve disentanglement. Our approach is a time-masking approach; however, unlike traditional random time-masking, we propose to mask all occurrences of certain speech units in an utterance. In random masking, phonetic units in the masked region could also be present in the unmasked region, making masking ineffective. In our approach, we ensure that the speaker encoder does not have access to some of the phoneme information in the input, reducing the reliance on input phonetic unit distribution. Our method is expected to maintain the capability of modeling speaker characteristics, relying on the assumption that not all phonetic units in the input utterance are needed to infer speaker identity, as the speaker can be fairly inferred from a short duration of speech where a limited number of phonetic units are present. 
\subsection{Discrete Speech Units}


A phonetic unit can be defined at various levels and groups such as phonemes, tri-phones, and even subphoneme units. However, such phonetic unit definitions require human knowledge about phonetics. Phonetic units should be determined by either manual labeling or an automatic approach such as speech recognition, which is computationally costly and has errors that would propagate. In this work, we opted for discrete units from self-supervised learning (SSL) models\cite{Hsu2021HuBERTSS,Chen2021WavLMLS} that are trained by masked unit prediction objective over speech frames and performed remarkably on automatic speech recognition (ASR) and many other downstream tasks. Discrete units are obtained by K-means clustering applied on intermediate layer representations of those SSL models. It is shown that the discrete units obtained by clustering the SSL features have high correlation and mutual information with phoneme classes \cite{Hsu2021HuBERTSS,Sicherman2023AnalysingDS}. We used a clustering model trained with pre-trained HuBERT \cite{Hsu2021HuBERTSS} layer 7 features to discretize input speech and determine corresponding unit classes for frames.

\subsection{Time-masking based on Discrete Speech Units}
In our approach (illustrated in Figure \ref{fig:framework}), we extract discrete units from input speech using HuBERT and a k-means model as $z = [z_1, z_2, ..., z_T]$ where $T$ is the number of frames. We get the set of discrete unit classes present in the utterance as $ \text{set}(z) = \{ z_k | z \}$ where $k$ is the discrete unit class index. We randomly select a portion of unit classes to mask in the set of discrete units and calculate timestamps of the corresponding frames classified as one of the selected unit classes for masking. We omit frames inside those timestamps in input speech features during training. We apply this masking only before feeding input features to the speaker encoder to avoid disrupting linguistic information needed by the content encoder, which could destabilize training by making the objective harder. By this approach, the speaker encoder is guaranteed not to have access to certain discrete units in the input speech, which are highly correlated with phoneme classes.
\vspace{-1mm}
\subsection{Masking Applied on VC Approaches}

To show the universality of our approach, we applied our masking approach to two different encoder-decoder based, zero-shot voice conversion frameworks, denoted as TriAAN-VC \cite{Park2023TriAANVCTA} and VQMIVC \cite{wang2021vqmivc}. Both methods rely on unsupervised and non-parallel training, where input speech utterance is encoded into linguistic and speaker representations by content and speaker encoders. The representations are then decoded back to reconstruct input speech. At inference, the voice is converted by decoding the linguistic representation from source speech and the speaker representation from a reference speech of the target speaker. This VC paradigm relies on the disentanglement of the speaker and linguistic features to maintain the source linguistic content while altering the speaker identity \cite{sisman2020overview}. Next, we elaborate on the specifics of the proposed masking approach for VC. 

\subsubsection{TriAAN-VC}


Triple Adaptive Attention Normalization VC (TriAAN-VC) \cite{Park2023TriAANVCTA} is an encoder-decoder based zero-shot VC framework. It has shown SOTA performance, especially in speaker similarity of the synthesized speech with the help of attention mechanisms introduced. It has two encoders for linguistic and speaker feature extraction followed by a bottleneck layer and decoder. It utilizes adaptive instance normalization (AdaIN) for disentanglement between linguistic and speaker features \cite{adain-vc,Huang2017ArbitraryST}. Furthermore, it is shown to improve speaker information extraction by incorporating attention while calculating instance normalization statistics utilizing fine-grained, frame-level details. 

The two encoders of TriAAN-VC consist of multiple blocks containing convolutional layers and instance normalization. The speaker encoder has an additional self-attention layer on the time dimension to calculate instance normalization statistics.

After encoding the features, a bottleneck layer consisting of a gated recurrent unit (GRU) and dual adaptive normalization (DuAN) is present. Following the bottleneck layer, the decoder consists of blocks containing convolutional layers, and the TriAAN module produces the mel spectrogram of the input utterance during training and the converted utterance during inference.

\textbf{Dual Adaptive Normalization:} 



In TriAAN-VC, DuAN calculates the weighted statistics of the speaker features for the stylization of content features after instance normalization. It is applied on both channel and time dimension separately hence it is called dual. The weights $\alpha$ for weighted statistics are calculated by the cross-attention between content features $x_c$ and speaker features $F_s$. The instance normalized content feature $IN(x_c)$ is then styled by weighted mean $M$ and standard deviation $S$ as $x_c' = IN(x_c)S+M$. During training, the input utterance is the same for the speaker and linguistic encoder. This suggests that cross-attention would easily get corresponding similar speaker features for each content feature during training. However, in inference, the reference speech from the target speaker most likely won't have the same linguistic and phonetic content as the source speech. This paradigm has an additional phonetic dependency on speaker features, which would suffer from overfitting and mismatch during inference. We believe our method would be especially effective in such cases.


\textbf{Loss:} TriAAN-VC uses L-1 reconstruction loss between predicted mel-spectrogram $\hat{y}$ and ground-truth mel-spectrogram $y$ as $loss = ||\hat{y} - y||_1$. Furthermore, they utilize siamese loss $siam=||\hat{y}_{siam} - y||_1$ where it is the L1-loss between ground truth $y$
predicted spectrogram $\hat{y}_{siam}$ from random time masking applied input before both encoders. They also include consistency loss $cons=||\hat{y}- \hat{y}_{siam}||_1$ which is the L1-loss between predicted mel-spectrogram from actual input and predicted spectrogram from random time masked input. The overall loss function becomes $L = (loss + siam)/2 + cons$. In our work, we only apply our masking when predicting $\hat{y}$. For the sake of training stability, we did not choose to apply our masking when predicting $\hat{y}_{siam}$ where the input is already masked by random time masking. In this work, the $loss$ and $cons$ terms in the overall loss promote reducing phonetic dependency while the $siam$ term promotes robustness for extracting content information.
\vspace{-2mm}
\subsubsection{VQMIVC}
Similarly, VQMIVC is an encoder-decoder based zero-shot VC framework. It utilizes vector-quantization (VQ) for content encoding and mutual information (MI) minimization between encoded features for disentanglement. It consists of two encoders $E_c,E_s$ for content and speaker identity respectively, a pitch extractor $E_p$, and a decoder $D$. It uses log-mel spectrogram as the input to the encoders and raw waveform for the pitch extractor.  The decoder reconstructs input mel-spectrogram from content $z_c$, speaker $z_s$, and pitch representations $z_p$ during training. They utilized vector-quantization with contrastive predictive coding (VQCPC) which is shown to learn speaker-independent linguistic units implicitly \cite{niekerk20b_interspeech,vqvc_wu20p_interspeech} as the content feature. To enforce the disentanglement further, the authors introduced mutual information estimation by utilizing vCLUB \cite{Cheng2020CLUBAC} between representation pairs. The authors include the estimated MI term in the overall loss function to minimize the common knowledge between the representations to improve the disentanglement explicitly.

However, both disentanglement approaches utilized in VQMIVC have limitations and can be further improved by our approach at the input side, where the access of the speaker encoder to the phonetic content of the speech is restricted. In this work, we apply our masking on the input log-mel spectrogram before feeding it to the speaker encoder $E_s$ for the speaker representation $z_s$ extraction during training. During inference, we don't apply masking. At the inference, the content $z_s$ and pitch $z_p$ representations are extracted from source speech, while speaker representation $z_s$ is extracted from a reference speech from a target speaker. The decoder $D$ produces a converted log-mel spectrogram. 
\vspace*{-3mm}

\section{Experimental Setup}
\vspace*{-3mm}
\subsection{Datasets}
We utilized the VCTK\cite{vctk} corpus for VC experiments, which consists of 109 speakers uttering 400 utterances each. We have followed the default data partition for TriAAN-VC\cite{Park2023TriAANVCTA} experiments and partitioned the dataset into a ratio of 60\%: 20\%: 20\% for training, validation, and testing respectively in terms of speakers. Similarly, for VQMIVC\cite{wang2021vqmivc} experiments, the dataset was partitioned into a training, validation, and testing ratio of 70\%: 10\%: 20\%. 
\vspace{-2mm}
\subsection{Training}
The audios are downsampled to 16 kHz sampling rate. 
For TriAAN-VC, we followed the default setting and CPC features from pre-trained SSL model\cite{Rivire2020UnsupervisedPT} is used as input features. For the f0 features, DIO algorithm\cite{dio} is used on raw waveforms for both methods. 
Log mel-spectogram features of 80 mel bins using a 25ms window and 10ms hop size are used as input features and output features for VQMIVC and only as output features for TriAAN-VC.   


We have used official implementations of TriAAN-VC\footnotemark[1] and VQMIVC\footnotemark[2] for training. We initialize default parameters for both frameworks according to \cite{wang2021vqmivc, Park2023TriAANVCTA} with no additional modifications other than proposed masking before the speaker encoder. 
We extract discrete units for input utterance using HuBERT-base and kmeans model \cite{Niekerk2021ACO} with $K=100$ offline. We apply masking to a fixed amount of the set of discrete units for a given utterance, where the masked units are selected randomly each time. We have experimented with different masked unit ratios of 10\%, 20\%, 30\% in different training settings.
The TriAAN-VC framework is trained using the Adam optimizer with a constant learning rate of 1e-4. A batch size of 64 is used to train to 400 epochs. The VQMIVC framework is trained using the Adam optimizer starting with a 15-epoch warmup which gradually increases the learning rate from 1e-6 to 1e-3. After 200 epochs, the learning rate is halved every 100 epochs until it reaches 500 epochs. The input utterances are segmented to a frame size of 128 for all of the training. Masking is applied to those segments and masked features are zero-padded at the end, to frame size. A batch size of 64 and 256 are utilized for TriAAN-VC and VQMIVC training, respectively. Parallel WaveGAN\cite{Yamamoto2019ParallelWA} vocoder pre-trained on the VCTK corpus is used to synthesize waveforms from predicted log mel-spectograms for both methods during inference. 

\footnotetext[1]{https://github.com/winddori2002/TriAAN-VC}
\footnotetext[2]{https://github.com/Wendison/VQMIVC}

\begin{table*}[!ht]
\centering
\caption{Objective Results for TriAAN-VC }
\label{table:triaan_obj}
\scalebox{0.8}{
\begin{tabular}{c|cc|cc|cc|c}
& \multicolumn{6}{c|}{Intelligibility} & \multicolumn{1}{c}{Similarity}\\
\hline
 & \multicolumn{2}{c|}{Conversion} & \multicolumn{2}{c|}{Resynth} & \multicolumn{2}{c|}{$\Delta$} & Conversion\\
 \hline
Methods & WER(\%) $\downarrow$ & CER(\%) $\downarrow$ & WER(\%) $\downarrow$ & CER(\%) $\downarrow$ & $\Delta$WER(\%) $\downarrow$ & $\Delta$CER(\%) $\downarrow$ & SECS $\uparrow$\\
\hline

TriAAN-VC & 22.81 & 12.37 & 8.74 & 3.37 & 14.07 & 9.00 & \textbf{70.77}\\
\hline
TriAAN-VC w/ random masking (10\%)& 26.41 & 14.03 & 11.41 & 4.43 & 15.00 & 9.60 & 70.05\\
TriAAN-VC w/ random masking (20\%) & 27.34 & 14.67 & 11.18 & 4.79 & 16.16 & 9.88 & 69.90\\
\hline
TriAAN-VC w/ our masking (10\%) & 15.89 & 7.66 & \textbf{7.97} & 3.08 & 7.92 & 4.58 & 69.49\\
TriAAN-VC w/ our masking (20\%) & 12.71 & 5.79 & 8.37 & \textbf{3.07} & 4.34 & 2.72 & 68.25\\
TriAAN-VC w/ our masking (30\%) & \textbf{11.40} & \textbf{5.14} & 8.20 & 3.21 & \textbf{3.20} & \textbf{1.93} & 67.48\\
\hline

\end{tabular}}
\end{table*}

\begin{table*}[!th]
\centering
\caption{Objective Results for VQMIVC}
\label{table:vqmi_obj}
\scalebox{0.8}{
\begin{tabular}{c|cc|cc|cc|c}
& \multicolumn{6}{c|}{Intelligibility} & \multicolumn{1}{c}{Similarity}\\
\hline
 & \multicolumn{2}{c|}{Conversion} & \multicolumn{2}{c|}{Resynth} & \multicolumn{2}{c|}{$\Delta$} & Conversion\\
 \hline
Methods & WER(\%) $\downarrow$ & CER(\%) $\downarrow$ & WER(\%) $\downarrow$ & CER(\%) $\downarrow$ & $\Delta$WER(\%) $\downarrow$ & $\Delta$CER(\%) $\downarrow$ & SECS $\uparrow$\\
\hline
VQMIVC & 21.90 & 11.06 & \textbf{19.54} & \textbf{9.96} & 2.36 & 1.10 & 63.35\\
\hline
VQMIVC w/ random masking(10\%) & 26.05 & 13.78 & 21.92 & 11.51 & 4.13 & 2.27 & 62.91\\
VQMIVC w/ random masking(20\%) & 28.38 & 15.23 & 23.90 & 12.55 & 4.48 & 2.68 & 62.60\\
\hline
VQMIVC w/ our masking (10\%) & \textbf{19.71} & \textbf{10.22} & 19.78 & 10.20 & \textbf{-0.07} & \textbf{0.02} & \textbf{63.44}\\
VQMIVC w/ our masking (20\%) & 21.70 & 11.38 & 20.76 & 10.81 & 0.94 & 0.57 & 62.35\\
\hline

\end{tabular}}
\end{table*}

\begin{table*}[!ht]
\centering
\caption{Subjective Results for TriAAN-VC with 95\% confidence interval }
\label{table:triaan_subj}
\scalebox{0.8}{
\begin{tabular}{c|c|c|c}
Method & Naturalness (MOS) $\uparrow$ & Intelligibility (MOS) $\uparrow$ & Similarity (SMOS) $\uparrow$\\
\hline
Ground-truth & 4.72 $\pm$ 0.11& 4.76 $\pm$ 0.10& - \\
\hline
TriAAN-VC & 3.39 $\pm$ 0.11& 3.34 $\pm$ 0.15  & 3.46 $\pm$ 0.11 \\
\hline
TriAAN-VC w/ our masking (10\%) & 3.68 $\pm$ 0.13& 3.68 $\pm$ 0.14 & 3.56 $\pm$ 0.11 \\
TriAAN-VC w/ our masking (20\%) & \textbf{3.99 $\pm$ 0.12} & \textbf{4.01 $\pm$ 0.12} & \textbf{3.65 $\pm$ 0.12} \\
\hline

\end{tabular}}
\end{table*}
\vspace*{-2mm}
\subsection{Evaluations}
 For the evaluations, we have performed VC for unseen speakers during the training. For TriAAN-VC, VC between pairs of 20 unseen speakers experimented. For the objective evaluations, 600 converted utterances are created. For VQMIVC, VC pairs are created from 20 unseen speakers totaling 720 converted utterances.
 
\textbf{Objective Evaluations:} In objective evaluations, we measured word error rate (WER) and character error rate (CER) when the synthesized speech is fed into a state-of-the-art automatic speech
recognition model\footnotemark[3] to measure intelligibility. To show the phonetic dependency and disentanglement, we have constructed a resynthesis scenario at inference where we don't use a reference utterance from the target speaker, but we use the source utterance itself as the reference utterance, just like in training. We report $\Delta$WER and $\Delta$CER between conversion and resynthesis, where the reference utterance has a different and same phonetic structure as the source utterance, respectively. We also measured speaker embedding cosine similarity (SECS) between embeddings from synthesized speech and ground truth utterances of the given speaker using the d-vector\footnotemark[4] speaker embeddings\cite{ge2e} to assess the similarity of converted utterance to the target speaker.
\footnotetext[3]{https://huggingface.co/facebook/wav2vec2-large-960h-lv60-self}
\footnotetext[4]{https://github.com/resemble-ai/Resemblyzer}

\textbf{Subjective Evaluations:}
We conducted subjective evaluations for TriAAN-VC with 15 participants. We have sampled 144 total utterances from conversion pairs of 16 speakers (8 male and 8 female) for the evaluations.  We have conducted mean opinion score\cite{sisman2020overview} (MOS) measurements for naturalness and intelligibility of the synthesized speech separately. We also asked participants to rate speaker similarity compared to the ground truth utterance from the target speaker for similarity mean opinion score\cite{pmlr-v162-casanova22a} (SMOS) measurements. 

\vspace{-3mm}
\section{Results}
\vspace*{-2mm}
In the experiments, we have applied the default setting for TriAAN-VC and VQMIVC as the baselines and applied our masking approach with different masked unit ratios. We also experimented with random time masking where a random segment consisting of consecutive frames is masked before feeding input to the speaker encoder. For random time masking we applied a fixed masked frame ratio with respect to the total number of frames in the input utterance. 


The objective results for TriAAN-VC can be seen in Table \ref{table:triaan_obj}. It is clear that the WER/CER difference between conversion and resynthesis scenarios is very high, indicating overfitting and being dependent on having the same or different linguistic, thus phonetic structure between source and reference. Our proposed masking approach improved WER/CER significantly where for both 20\% and 30\% masking the relative improvement is around 50\%. Furthermore, the difference between the conversion and resynthesis scenarios is also much smaller with our proposed masking. These results indicate the very effectiveness of our approach to remedy the dependence on phonetic structure and overfitting of the attention-based baseline. The WER/CER gets better with increasing masked unit ratio but at the same time, the speaker similarity decreases. We believe with increased masking, speaker information extraction becomes less reliable thus we opted for 20\% masking where the loss in speaker similarity is negligible compared to the gain in intelligibility. Random time masking decreases the intelligibility significantly, which does not help with the disentanglement and possibly compromises the training by introducing artifacts in the input utterance. This supports our motivation to mask all occurrences of a speech unit to be effective rather than random masking.

In the VQMIVC objective results, the difference in conversion/resynthesis scenario is smaller compared to TriAAN-VC. This indicates that attention-based methods suffer more from dependency on the phonetic structure of the input utterance. Similarly, our proposed method improves WER/CER in VQMIVC both for 10\% and 20\% masked unit cases. Furthermore, the disentanglement is also improved significantly as the $\Delta$WER/$\Delta$CER is much smaller in proposed cases, it even becomes negative in 10\% masking scenario.

The subjective results can be seen in Table \ref{table:triaan_subj}. For intelligibility proposed method significantly performed better than the baseline for both 10\% and 20\% masking, the latter having an impressive subjective score. The same pattern is observed in naturalness where both masking scenarios are significantly better than baseline where 20\% masking has a very high naturalness score. For the speaker similarity, all the methods have similar results just like in the similarity objective metric. However, the proposed 20\% sampling has a slightly better result where we believe significantly improved naturalness might affected the listeners. Subjective results show that our proposed approach improves intelligibility and naturalness remarkably without compromising speaker similarity.

\vspace*{-3mm}
\section{Conclusion}
\vspace*{-2mm}
In this study, we proposed a straightforward yet very effective approach to reduce the phonetic dependency of speaker features and improve disentanglement in zero-shot, encoder-decoder based VC architectures. We applied our approach to multiple VC frameworks and showed that our approach has improved intelligibility and disentanglement in different VC methods suggesting universality.
We have shown that attention-based VC frameworks especially suffer from this dependency because of the connection between content and speaker features in the training paradigm, which has an overfitting tendency. Our method is very effective in this paradigm, with remarkable improvements in intelligibility and naturalness resulting in a high-quality SOTA VC. In future work, we plan to apply our method to other VC frameworks and investigate the approaches where the phonetic dependency can be remedied more effectively by an objective in the loss function. 
\vspace*{-3mm}
\section{ACKNOWLEDGMENTS}
\vspace*{-2mm}
This work was funded by NSF CAREER award IIS-2338979.
\bibliographystyle{IEEEbib}
\bibliography{strings,refs}
\end{document}